# Structural and Transport Properties of Thin InAs Layers Grown on In$_x$Al$_{1-x}$As Metamorphic Buffers

Giulio Senesi, Katarzyna Skibinska, Alessandro Paghi, Gaurav Shukla, Francesco Giazotto, Fabio Beltram, Stefan Heun, and Lucia Sorba*

*Istituto Nanoscienze-CNR and Scuola Normale Superiore, Piazza San Silvestro 12, 56127 Pisa, Italy*

\*      Correspondence: lucia.sorba@nano.cnr.it

Indium Arsenide is a III–V semiconductor with low electron effective mass, a small band gap, strong spin–orbit coupling, and a large g-factor. These properties and its surface Fermi level pinned in the conduction band make InAs a good candidate for developing superconducting solid-state quantum devices. Here, we report the epitaxial growth of very thin InAs layers with thicknesses ranging from 12.5 nm to 500 nm grown by Molecular Beam Epitaxy on In$_x$Al$_{1-x}$As metamorphic buffers. Differently than InAs substrates, these buffers have the advantage of being insulating at cryogenic temperatures, which allows for multiple device operations on the same wafer and thus making the approach scalable. The structural properties of the InAs layers were investigated by high-resolution X-ray diffraction, demonstrating the high crystal quality of the InAs layers. Furthermore, their transport properties, such as total and sheet carrier concentration, sheet resistance, and carrier mobility, were measured in the van der Pauw configuration at room temperature. A simple conduction model was employed to quantify the surface, bulk, and interface contributions to the overall carrier concentration and mobility.

## 1. Introduction

Indium Arsenide is a III–V semiconductor with interesting properties such as low electron effective mass, a small band gap, strong spin–orbit coupling, a large g-factor, and a surface Fermi level pinned at the conduction band edge. Such properties make InAs a good candidate for developing superconductor–semiconductor solid-state quantum devices [1].

Growing high-quality InAs films is still an open issue. Commercially available substrates, such as InAs, GaAs, InP, and GaSb, can be employed to grow InAs layers. The lattice-matched homoepitaxy growth of InAs on InAs substrates leads to high-quality material. Still, InAs substrates are conductive and not suitable for the realization of insulating planar superconductor–semiconductor quantum devices. On the other hand, GaSb substrates are only 0.06% lattice mismatched with InAs, but they are quite expensive. InP and GaAs substrates are 3.2% and 7.2% lattice mismatched, respectively. Therefore, the growth of high crystal quality InAs films on these substrates is quite challenging.

Vacuum deposition of InAs films on GaAs substrates was carried out during the 1960s. These studies showed that a growth temperature of 600 °C and an Arsenic overpressure were necessary to realize good-quality InAs films [2–4]. Later, by employing the Molecular Beam Epitaxy (MBE) technique, M. Yano et al. [5] achieved higher mobilities growing 1 µm thick InAs films on GaAs substrates at a temperature of 480 °C and with a V/III ratio of 10–12. Instead, they found that InAs films grown at very low temperatures (less than 350 °C) were polycrystalline, showing Debye–Scherrer rings and spots in the reflection high-energy electron diffraction (RHEED) patterns. At the same time, for too-high growth temperatures, Indium droplets were observed on the sample surface [5]. B. T. Meggitt et al. [6] found the optimal growth temperature of 500 °C for InAs film thickness in the 0.5–2.0 µm range. Further improvement in InAs epilayers grown by MBE on GaAs has been obtained by introducing an In$_{0.8}$Ga$_{0.2}$As/GaAs strained superlattice (SLS) before the InAs growth. The SLS makes a superior interfacial transition at the layer/interface and reduces the film defect density [7]. In the growth temperature range from 470 °C to 510 °C and with a V/III ratio around three, the optimal growth was achieved by P. D. Wang et al. [8], who reported an electron mobility as high as $3.5 \times 10^4$ cm$^2$/(V·s) for a 2.4 μm InAs thickness, but the surface morphology was relatively poor. Moreover, it was found that the electron mobility increases with increasing layer thickness, while the bulk carrier concentration decreases [6,7]. In addition to MBE, Metal–Organic Chemical Vapor Deposition (MOCVD) is another widely used technique for the growth of III-V semiconductors. In recent years, MOCVD has been used for the growth of high-efficiency light-emitting diodes [9], quantum cascade lasers [10], and in general also for the integration of III-V devices in silicon technology [11]. Moreover, starting from the 1970s [12], MOCVD has been used to grow thick InAs films on GaAs substrates. In particular, the same thickness dependence of InAs transport properties reported for MBE-grown samples has been also found for InAs layers grown by MOCVD on GaAs [13].

Hall effect measurements on InAs films have been performed to investigate the temperature dependence of carrier concentration and mobility. An absence of carrier freeze-out in temperature-dependent Hall measurements was generally found. The lack of freeze-out was attributed to surface and interface conduction channels present in the InAs layer [6,14]. Moreover, S. Kalem et al. [7] investigated the temperature dependence of electron mobility in 5.2 μm and 6.2 μm thick InAs films. They found that temperature-dependent electron mobility goes as $T^{-1.1}$ instead of $T^{-1.5}$, expected for bulk InAs (for T > 100 K). This result suggests that different scattering mechanisms other than impurity and optical phonon scattering are present in the InAs epilayer. Good agreement with the experimental data was found by introducing dislocation scattering screened by free electrons [15] for the thinner samples. This result is also consistent with previous work [16] that correlates the fast drop in mobility to dislocation density and not only to the surface scattering mechanism.

This paper investigates the MBE growth, crystal quality, and transport properties of very thin InAs layers with thicknesses ranging from 12.5 nm to 500 nm epitaxially grown on In$_x$Al$_{1-x}$As metamorphic buffers. Transport properties, such as total and sheet carrier concentration, sheet resistance, and carrier mobility, were measured in the van der Pauw configuration at room temperature. A simple conduction model was employed to quantify the surface, bulk, and interface contributions to the overall carrier concentration and mobility.

## 2. Materials and Methods

The InAs thin layers presented in this paper were grown on semi-insulating GaAs(100) substrates (resistivity higher than $2.37 \times 10^7$ Ω·cm and carrier concentration lower than $1.3 \times 10^8$ cm$^{-3}$) using solid-source MBE in a Riber compact 21 DZ system (Riber, Paris, France). Schemes of the layer structures are shown in Figure 1. For the samples shown in Figure 1a, starting from the GaAs substrate, the structure consists of a 50 nm GaAs layer, a 100 nm GaAs/AlGaAs superlattice (SL), a 50 nm GaAs layer, a 1250 nm step-graded In$_x$Al$_{1-x}$As metamorphic buffer layer with x increasing from 0.15 to 0.81, a 400 nm overshoot (OS) layer with an Indium concentration x = 0.84, and InAs epilayers with a thickness ranging from 12.5 nm to 500 nm. The In$_x$Al$_{1-x}$As metamorphic buffer consists of two regions with different misfit gradients df/dt (5.1% μm$^{-1}$ and 3.1% μm$^{-1}$, respectively). The first region consists of thirteen 50 nm steps with x increasing from 0.15 to 0.58, while the second region comprises twelve 50 nm steps with x rising from 0.58 to 0.81. The OS layer helps to overcome the residual strain of the grown metamorphic buffer. The growth procedure was adapted from [17]. The group V/III beam flux ratio of 7 was kept constant during the growth by adjusting the Arsenic flux. The buffer and the InAs layers were grown at optimized substrate temperatures of 340 °C and 500 °C, respectively, measured by a WRe 5/26 thermocouple. Samples with different InAs thicknesses t of 12.5 nm, 25 nm, 50 nm, 75 nm, 100 nm, 300 nm, and 500 nm are labeled A, B, C, D, E, F, and G, respectively. A 5 nm capping layer of undoped In$_{0.84}$Al$_{0.16}$As was grown above 50 nm (sample H) and 100 nm (sample I) thick InAs layers (Figure 1b).

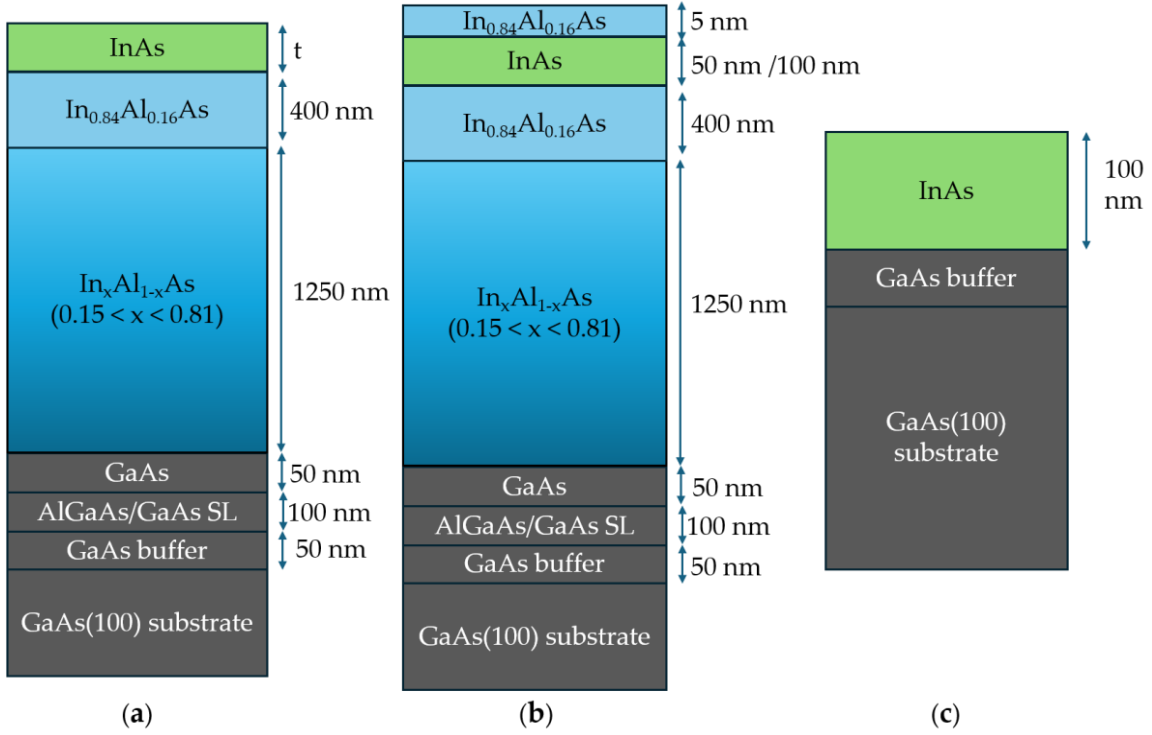

**Figure 1.** (**a**) Sample structure of InAs layer grown on $In_xAl_{1-x}As$ metamorphic buffer. (**b**) Sample structure of InAs layer grown on $In_xAl_{1-x}As$ metamorphic buffer with capping layer. (**c**) Sample structure of InAs layer grown on GaAs substrate.

Nomarski optical microscopy was used to evaluate the surface morphology of each grown sample. For the InAs on metamorphic buffers and on the GaAs substrate samples, μm sized defects are present. These defects are not oval defects, but they are probably related to the lattice-mismatched heteroepitaxy. Their number depends on the growth conditions (III/V ratio and growth temperature). To estimate the surface defect density, five random images of the surface are taken in different regions from each sample (for some examples, see the Supplementary Materials). For each image, the surface defect density is estimated as the number of defects over the image area. The reported surface defect density and its error are the mean and standard deviation of the five different images, respectively. For all the InAs epilayers grown on metamorphic buffers, the average surface defect density is $10 \pm 2$ $mm^{-2}$.

As a reference, 100 nm thick InAs layers were deposited directly on GaAs (Figure 1c). The sample, called J, with InAs grown at the same conditions as the samples grown on a metamorphic buffer (growth temperature of 500 °C and a group V/III beam flux ratio of 7), showed a milky surface with a surface defect density greater than $5 \times 10^5$ $mm^{-2}$. Therefore, the growth temperature was increased to 560 °C and the Arsenic flux was adjusted to a group V/III beam flux ratio of 10 (sample K), which resulted in a shiny mirror-like InAs layer on GaAs with a surface defect density of $9 \pm 7$ $mm^{-2}$.

High-resolution X-ray diffraction (HR-XRD) measurements were performed to investigate the perpendicular residual strain in OS and the InAs layers using a QC3 Bruker (Bruker, Migdal Ha'Emek, Israel) diffractometer. The X-ray tube consists of a tungsten filament and a copper anode, operating at a voltage of 40 kV and a current of 40 mA. The X-ray optics consist of a parabolic graded multilayer mirror and a channel-cut collimator with a 2-bounce Ge(004) crystal. The parabolic mirror parallelizes the beam with a divergence of ~0.1°, while the 2-bounce Ge(004) monochromator selects the Cu $K_\alpha$ wavelength (1.54059 Å). Symmetric (004) ω-2θ scans were performed to calculate the out-of-plane lattice constants of the lattice-mismatched epilayers. The crystal quality of the InAs layers was evaluated using the full width at half maximum (FWHM) of the InAs peak in the spectra. A reflection high-energy electron diffraction (RHEED) system from STAIB Instruments (Munich, Germany) was used to monitor surface morphology during the growth of epilayers.

Transport measurements of all samples were performed at 300 K using the van der Pauw method. InAs samples were cut into squares (7 × 7 mm$^2$) and sonicated (Transonic, T310/H) in ACE and IPA for 5 min each. Before applying resist, the surface structure was cleaned in O$_2$ plasma for 60 s at 100% power. LOR 3A resist was spin-coated at 4 krpm for 60 s and soft-baked at 175 °C. Then, a layer of S1805 positive photoresist was spin-coated at 5 krpm for 60 s and soft-baked at 115 °C. The van der Pauw contact area (500 × 500 μm$^2$ each square) was exposed using an M3 laser writer. The pattern was developed in MF319 for 2 min, followed by DI water for 15 s, before drying by N$_2$ flux. Samples were passivated in an (NH$_4$)$_2$S$_X$ solution (290 mM (NH$_4$)$_2$S and 300 mM S in deionized water) at 45 °C for 60 s. A 100 nm thick Al layer was deposited in a thermal evaporator at a base pressure of 5 × 10$^{-6}$ mbar. The lift-off process was performed in AR600-71 remover at 80 °C for 5 min followed by a 30 s IPA dip. Then, the sample was mounted on the chip carrier using ARP679.04 EBL resist and bonded with Al wire.

## 3. Results and Discussion

### 3.1. Structural Properties

For all grown samples, symmetric (004) scans were measured in triple-axis mode along 0° and 180° azimuths. Figure 2a shows the ω-2θ scans of the series of samples with increasing InAs layer thickness, grown on In$_x$Al$_{1-x}$As metamorphic buffers, measured under the azimuthal angle Φ = 0°. For all scans, the integration time was 1 s, and the step size was five arcsec. Symmetric (004) scans of the InAs grown directly on the GaAs substrate (samples J and K) are shown in Figure 2b. In this case, an integration time of 1 s was used for the InAs peak. For the rest of the scan range, this time was 0.1 s.

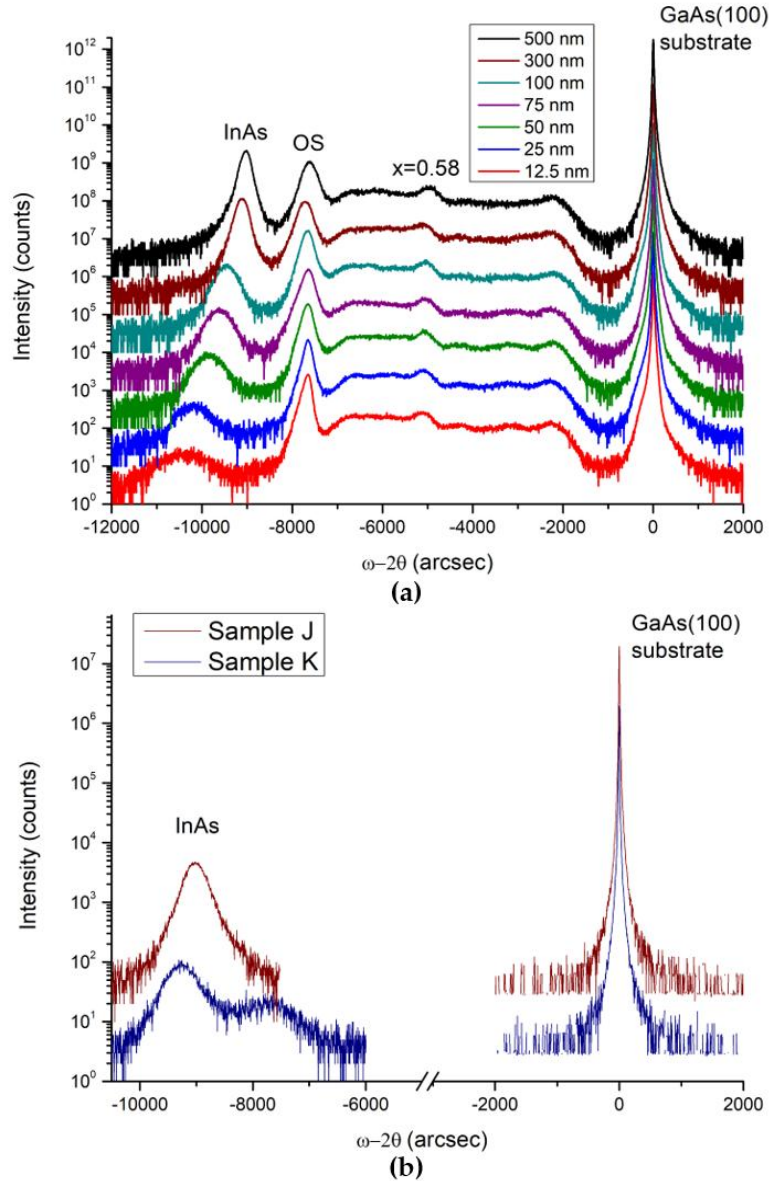

**Figure 2.** Symmetric (004) ω-2θ scans obtained along the 0° azimuth from (**a**) a series of samples with different InAs thicknesses grown on InAlAs metamorphic buffers and (**b**) 100 nm thick InAs layers grown directly on a GaAs substrate at a sample temperature of 500 °C (sample J) and 560 °C (sample K).

Three prominent peaks are present in all spectra (Figure 2a): the substrate, the OS, and the InAs layer peaks. The most intense peak corresponds to the GaAs substrate. The broad features in the range from −7000 arcsec to −1500 arcsec originate from the $In_xAl_{1-x}As$ metamorphic buffer. A small peak corresponding to an Indium content of 0.58 appears around −5000 arcsec. This peak is associated with a slightly thicker $In_{0.58}Al_{0.42}As$ layer with an In composition that corresponds to the change in the grading slope of the metamorphic buffer. A similar position of this peak and the OS peak in all scans indicates the reproducibility of the growth conditions. The InAs peak shifts to the right with increasing thickness due to different strain relaxation in these layers. Figure 2b shows symmetric (004) ω-2θ scans of the InAs layers grown directly on the GaAs substrates. Only two peaks are observed in the spectra, corresponding to the GaAs substrate and InAs layer. The position of the InAs peak varies between samples, which suggests different strain states for the InAs layers. A clear shoulder is present on the right side of the InAs peak in the spectrum of sample K. Since we can rule out the presence of tilting (for details, see the Supplementary Materials), we suggest that it is caused by the diffusion of Ga into the InAs film [18], resulting in a graded InGaAs layer formed immediately after the GaAs interface. This phenomenon is favored in sample K due to the higher growth temperature (560 °C) compared to sample J (500 °C). Moreover, the RHEED pattern of sample J, a few seconds after the opening of the Indium shutter, was spotty. No lines were present even after the 100 nm thick InAs layer was grown. This suggests that the growth mode of this sample is primarily 3D, which is a common feature in InAs/GaAs growth [19]. Instead, for sample K, no spots were present in the RHEED pattern, while well-visible reconstruction streaks in the 2× and 4× directions were present during all InAs growth.

These ω-2θ scans allowed the determination of the residual perpendicular strain in OS and InAs layers. The Pseudo-Voigt function was implemented to extract the peak positions of the GaAs substrate, OS, and InAs peaks. The Bragg angle of each peak is derived from the fitting procedure, and so is the out-of-plane lattice constant $a_\perp$. The following formulas were used:

$$\varepsilon_{OS,\perp} = \frac{a_{OS,\perp} - a_{0.84}}{a_{0.84}} \qquad (1)$$

$$\varepsilon_{InAs,\perp} = \frac{a_{InAs,\perp} - a_{InAs}}{a_{InAs}} \qquad (2)$$

where $a_{OS,\perp}$ and $a_{InAs,\perp}$ are the measured lattice constants, while $a_{0.84}$ and $a_{InAs}$ are the relaxed lattice constants of $In_{0.84}Al_{0.16}As$ and InAs, respectively. In this analysis, the overshoot In composition is considered the same for all samples and equal to the nominal value. This is a reasonable assumption, considering the careful In composition calibration by XRD (we grew two calibration samples, which gave a consistent OS composition of 0.84 ± 0.01). The error on the residual perpendicular strain was estimated from the error in the peak position. It was calculated as the sum of the squares of the scan step size, the average peak shift for different azimuthal directions, and the peak fitting error.

The relaxation (*Rel.*) was also determined for the InAs layers grown on a metamorphic buffer:

$$Rel. = \frac{a_{InAs,\parallel} - a_{0.84}}{a_{InAs} - a_{0.84}} \qquad (3)$$

However, since $a_{InAs,\parallel}$ cannot be measured from a symmetric scan, supposing tetragonal distortion, its value was calculated using

$$\left(\frac{a_{InAs} - a_{0.84}}{a_{0.84}}\right) = \frac{1-\nu}{1+\nu}\left(\frac{a_{InAs,\perp} - a_{0.84}}{a_{0.84}}\right) + \frac{2\nu}{1-\nu}\left(\frac{a_{InAs,\parallel} - a_{0.84}}{a_{0.84}}\right) \qquad (4)$$

where $\nu = 0.35$ is the Poisson ratio of InAs [20].

Relaxation of the InAs layers grown on a GaAs substrate was also determined. The value of the relaxation was calculated for the InAs layer with respect to $In_{0.84}Al_{0.16}As$ (samples A–I) or GaAs (samples J and K). All data are summarized in Table 1. The residual perpendicular strain of the InAs layers as a function of InAs thickness is shown in Figure 3.

**Table 1.** Summary of $\varepsilon_{OS,\perp}$, $\varepsilon_{InAs,\perp}$, and *Rel.* The *Rel.* of the InAs layers is calculated with respect to InAlAs with a 0.84 Indium fraction (samples A to I), while for samples J and K it is calculated with respect to GaAs.

| Sample | t (nm) | $\varepsilon_{OS,\perp}$ (×10$^{-3}$) | $\varepsilon_{InAs,\perp}$ (×10$^{-3}$) | *Rel.* (%) |
|:---:|:---:|:---:|:---:|:---:|
| A | 12.5 | 0.8 ± 0.2 | 13.1 ± 1.4 | ~0 |
| B | 25 | 0.8 ± 0.2 | 11.1 ± 1.6 | 3 |
| C | 50 | 0.9 ± 0.2 | 8.3 ± 1.2 | 27 |
| D | 75 | 0.8 ± 0.2 | 6.5 ± 0.5 | 43 |
| E | 100 | 0.9 ± 0.2 | 5.1 ± 0.4 | 55 |
| F | 300 | 1.3 ± 0.2 | 2.2 ± 0.3 | 81 |
| G | 500 | 0.5 ± 0.2 | 1.5 ± 0.3 | 90 |
| H | 50 | 0.9 ± 0.2 | 8.5 ± 0.9 | 25 |
| I | 100 | 0.8 ± 0.2 | 5.0 ± 0.4 | 56 |
| J | 100 | no OS | 1.5 ± 0.4 | 98 |
| K | 100 | no OS | 3.5 ± 0.6 | 95 |

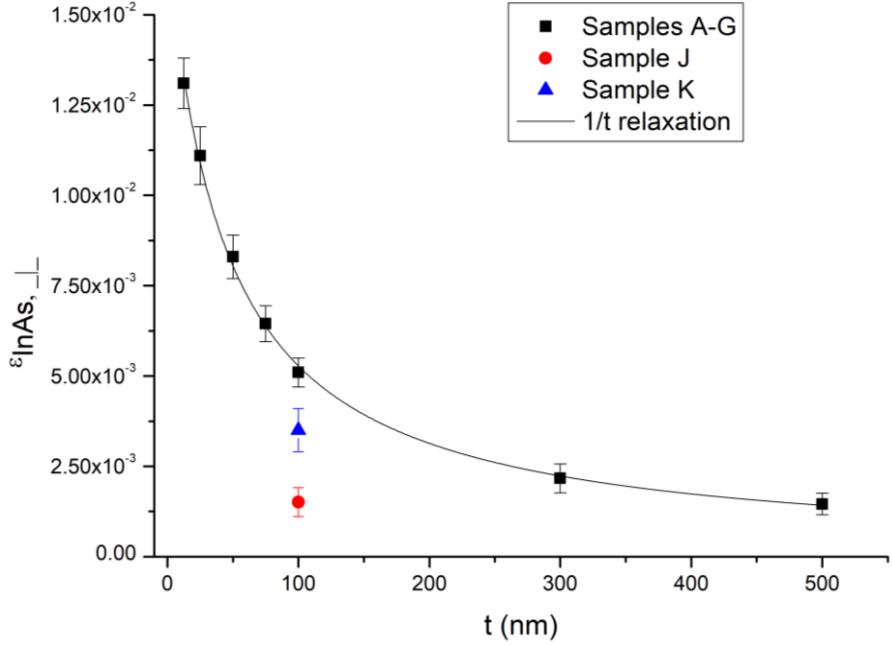

**Figure 3.** Perpendicular strain of the InAs layers as a function of InAs thickness. The black squares correspond to samples A–G, while the red dot and blue triangle correspond to samples J and K, respectively. A fit function that scales 1/t is reported as a solid black line.

The values of $\varepsilon_{OS\perp}$ are comparable within the experimental error, with samples F and G as slight outliers (see Table 1). However, right after sample G, we grew a calibration sample that gave x = 0.83, consistent with the original calibration (x = 0.84) within the error (x = 0.01). The more the InAs peak is shifted to the left (Figure 2a), the larger the values of the out-of-plane lattice constant and perpendicular strain. The critical thickness $t_c$ of InAs on this metamorphic buffer must be slightly larger than the thickness of strained InAs quantum wells grown on similar metamorphic buffer layers, which is 7 nm [17,21]. This means that a critical thickness of ≃10 nm is expected. For t = 12.5 nm, the InAs thickness is of the order of $t_c$, so almost no relaxation is expected. This is consistent with the value of relaxation given in Table 1. For all the other samples, the InAs epilayer thickness is larger than the critical thickness $t_c$, and dislocations are nucleated near the overshoot layer. Plastic relaxation occurs when the grown layer thickness exceeds $t_c$, and the thicker the layers, the more the strain is relaxed [22]. Relaxation is taking place for samples B, C, D, E, and F. For 500 nm (sample G), the InAs layer is nearly completely relaxed (90%) with respect to $In_{0.84}Al_{0.16}As$, as reported in Table 1. The relaxation of the InAs layers agrees with the geometrical model for strain relaxation proposed in Ref. [22], for which the strain scales as 1/t with the epilayer thickness. In Figure 3, a fit function that scales as 1/t is reported as a solid black line. We note that the strain in sample J, grown at 500 °C directly on a GaAs substrate, is within the experimental error comparable with the strain of the 500 nm thick InAs layer on the metamorphic buffer (sample G). Sample K, grown at 560 °C, is less relaxed than sample J. The fact that samples J and K are more relaxed than the 100 nm thick InAs sample on the metamorphic buffer agrees with our expectation, since the critical thickness of the InAs on GaAs is ≃3 ML, which is about 10 times smaller than the expected critical thickness of the InAs on the metamorphic buffer. For this reason, for the InAs directly grown on GaAs, most of the relaxation takes place for a much smaller layer thickness. The different strain values reported for samples J and K are consistent with the presence of a shoulder in the XRD scan for sample K due to Ga diffusion into the InAs layer, as reported in Figure 2b.

To determine the InAs crystal quality, the InAs peak of each sample is fitted with a Pseudo-Voigt function to extract the FWHM of the peak. The dependence of FWHM on the thickness of the InAs layers is shown in Figure 4. Samples grown directly on a GaAs substrate are also reported in Figure 4.

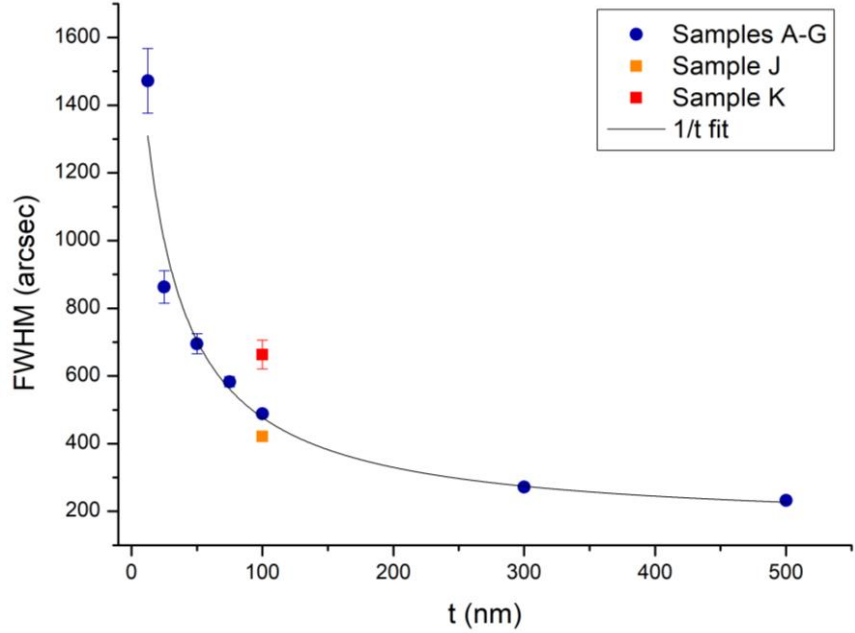

**Figure 4.** FWHM of the InAs peak as a function of InAs film thickness of samples grown on metamorphic buffers (samples A–G, dots) and on GaAs substrates (samples J and K, squares). The continuous line is the result of the fitting of the FWHM versus 1/t.

We notice that the FWHM of the InAs peak increases with decreasing InAs thickness t. Generally, we expect that the thickness dependence of the FWHM of the InAs peak could be caused by two different effects: defect broadening (caused by dislocations) and thickness broadening. For thickness broadening, we would expect a 1/t dependence [23,24], as experimentally observed. On the other hand, since the InAs layer of sample A is not relaxed, i.e., pseudomorphic (as reported in Table 1), we expect it to be less defective compared to thicker InAs films on the metamorphic buffer, which are more relaxed and therefore contain more dislocations. Thus, defect broadening appears to be inconsistent with the experimental data. Therefore, we conclude that the leading effect that causes the behavior reported in Figure 4 is a thickness broadening of the InAs peak. We also note that the FWHMs of samples J and K are not the same. While the InAs peak of sample J and sample E have similar FWHMs, the surface of sample J is full of defects and unsuitable for device fabrication. On the other hand, sample K has a surface defect density comparable to samples A–G, but the FWHM of the InAs layer is $664 \pm 42$ arcsec. This value is nearly 180 arcsec larger than the FWHM of sample E. This means that for the growth of a sample with an almost defect-free surface, better crystal quality is obtained growing the 100 nm thick InAs layer on top of an InAlAs metamorphic buffer.

*3.2. Transport Properties*

Hall effect measurements in the van der Pauw geometry were performed at 300 K to determine the value of volume carrier concentration n, sheet carrier concentration $N_s$, sheet resistance $R_s$, and electron mobility $\mu_H$. The error reported for carrier concentration and sheet resistance is estimated by performing the Hall effect measurements at least two times on the same sample. The scatter between the two measured values is about 10%, and we indicate this as the experimental error. Instead, the relative error on the electron mobility is calculated as the sum of squares of the relative error on sheet resistance and carrier concentration, resulting in a 14% relative error.

Sheet carrier concentration ($N_s$) and volume carrier concentration (n) are reported in Figures 5 and 6, respectively. The values for samples H, I, J, and K are summarized in Table 2. Sheet carrier concentration ($N_s$) is calculated as n multiplied by the InAs thickness t. We note that the sheet carrier concentration only depends on the measured Hall voltage, and no assumptions on the InAs layer thickness are needed.

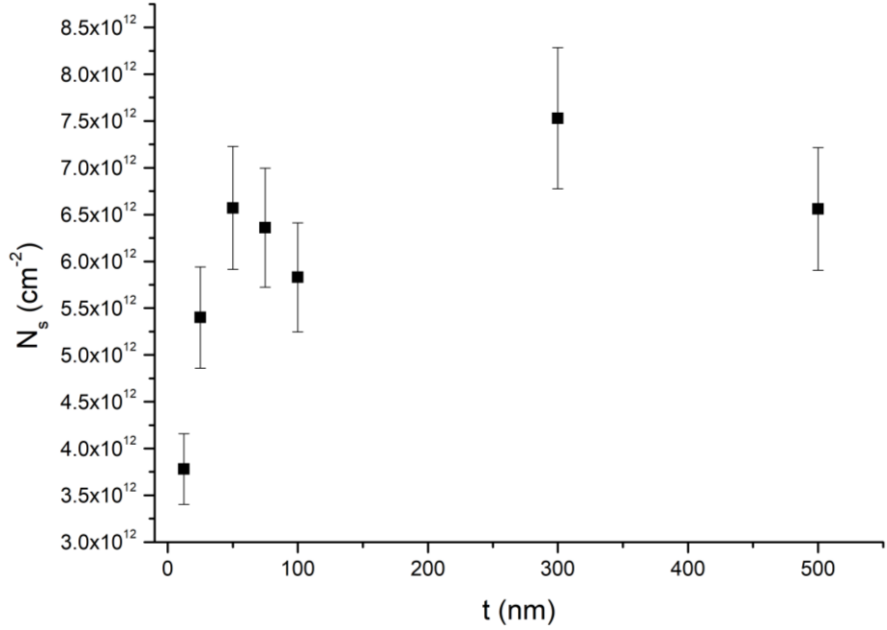

**Figure 5.** Sheet carrier concentration $N_s$ of samples A–G versus InAs thickness t measured in van der Pauw geometry at 300 K.

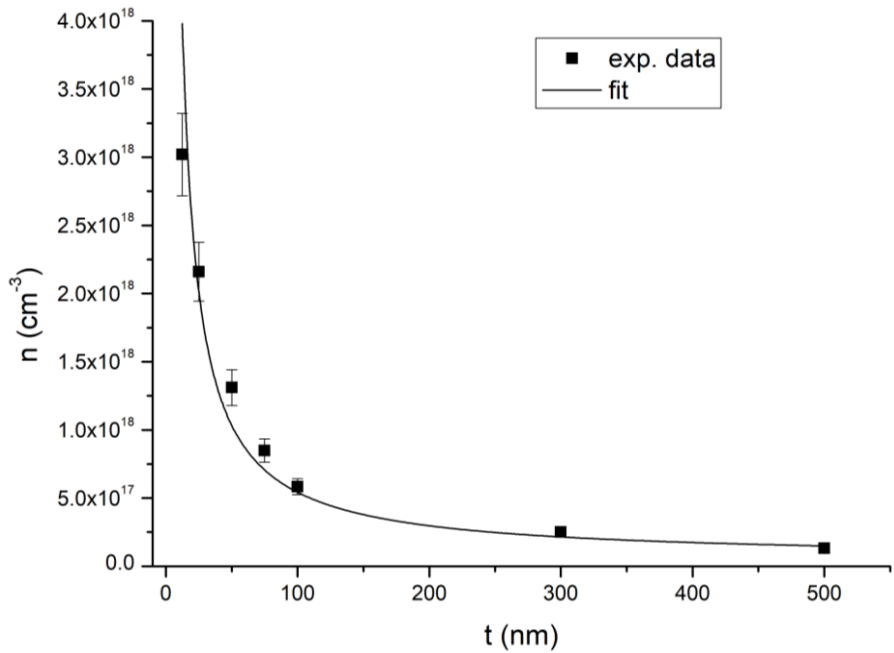

**Figure 6.** The volume carrier concentration n of samples A–G versus InAs thickness t, measured at 300 K using the van der Pauw geometry. The continuous line is the result of the fitting procedure to Equation (8).

**Table 2.** Summary of sheet carrier concentration ($N_s$) and volume carrier concentration (n) of samples H-K at 300 K.

| Sample | t (nm) | $N_s$ (×$10^{12}$ cm$^{-2}$) | n (×$10^{17}$ cm$^{-3}$) |
|---|---|---|---|
| H | 50 | 3.8 ± 0.4 | 7.6 ± 0.8 |
| I | 100 | 4.2 ± 0.4 | 4.2 ± 0.4 |
| J | 100 | 6.8 ± 0.7 | 6.8 ± 0.7 |
| K | 100 | 9.1 ± 0.9 | 9.1 ± 0.9 |

As we can see in Figure 5, for values of t less than 50 nm, the sheet carrier concentration decreases with decreasing t. In comparison, for values of t greater than 50 nm, it is essentially constant within the experimental error. Indeed, the average sheet carrier concentration of samples (C-G) is (6.6 ± 0.6) × $10^{12}$ cm$^{-2}$, which is comparable within the error with the sheet carrier concentration of each sample (C–G). Volume carrier concentration presents a clear dependence on InAs thickness, as shown in Figure 6. The thicker the InAs layer, the lower the total carrier concentration.

Table 2 shows that the sheet carrier concentration of samples J and K is greater than the sheet carrier concentration of 100 nm InAs on the metamorphic buffer sample ($N_s$ = (5.8 ± 0.6) × $10^{12}$ cm$^{-2}$). Furthermore, we compare the sheet carrier concentration of the capped samples H and I to the values of the sheet carrier concentration of the 50 nm thick and 100 nm thick uncapped InAs on the metamorphic buffer samples. For t = 50 and 100 nm of uncapped samples (sample C and E), $N_s$ is equal to (6.6 ± 0.7) × $10^{12}$ cm$^{-2}$ and (5.8 ± 0.6) × $10^{12}$ cm$^{-2}$, respectively. Instead, for the 50 nm thick (sample H) and 100 nm thick (sample I) InAs capped samples, $N_s$ = (3.8 ± 0.4) × $10^{12}$ cm$^{-2}$ and (4.2 ± 0.4) × $10^{12}$ cm$^{-2}$, respectively. This means that, for samples H and I, $N_s$ is (2.8 ± 0.8) × $10^{12}$ cm$^{-2}$ and (1.6 ± 0.7) × $10^{12}$ cm$^{-2}$ lower than the sheet carrier concentration of the 50 and 100 nm thick InAs uncapped samples, respectively.

Figure 7 reports the sheet resistance $R_s$ of the samples with InAs layers grown on metamorphic buffers as a function of InAs thickness t. As shown in Figure 7, the sheet resistance decreases with increasing the InAs thickness. The values for samples H, I, J, and K are summarized in Table 3.

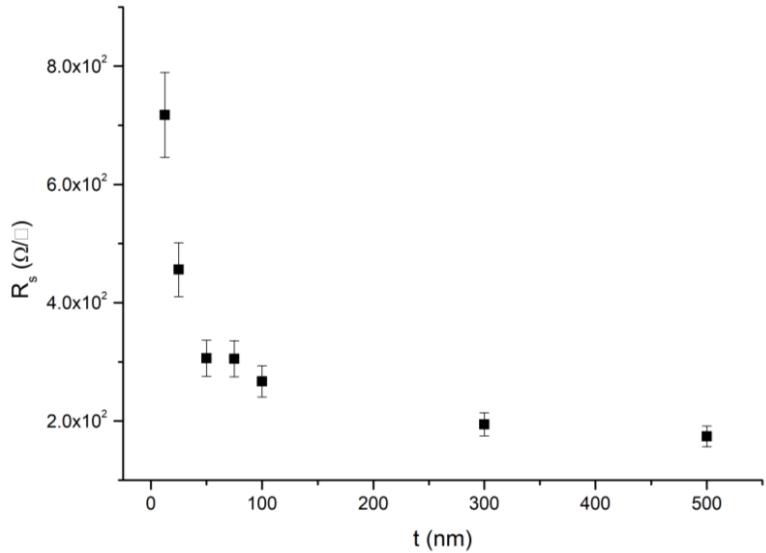

**Figure 7.** Sheet resistance $R_s$ of samples A–G versus InAs thickness t measured in van der Pauw geometry at 300 K.

For samples J and K, $R_s$ is greater than the value of the 100 nm thick InAs grown on the metamorphic buffer. Instead, the sheet resistance values measured for sample H and I are nearly comparable within the error with the sheet resistance values of the 50 nm and 100 nm thick InAs layers grown on the metamorphic buffer, respectively.

Figure 8 reports the Hall electron mobility for samples A–G as a function of InAs thickness. Table 3 summarizes the values of Hall mobility for samples H to K. As we can see in Figure 8, for the samples grown on metamorphic buffers, increasing the InAs thickness also increases the Hall mobility. Moreover, in samples J and K, the Hall mobility is almost a factor two lower with respect to the InAs grown on the metamorphic buffers for the same InAs thickness.

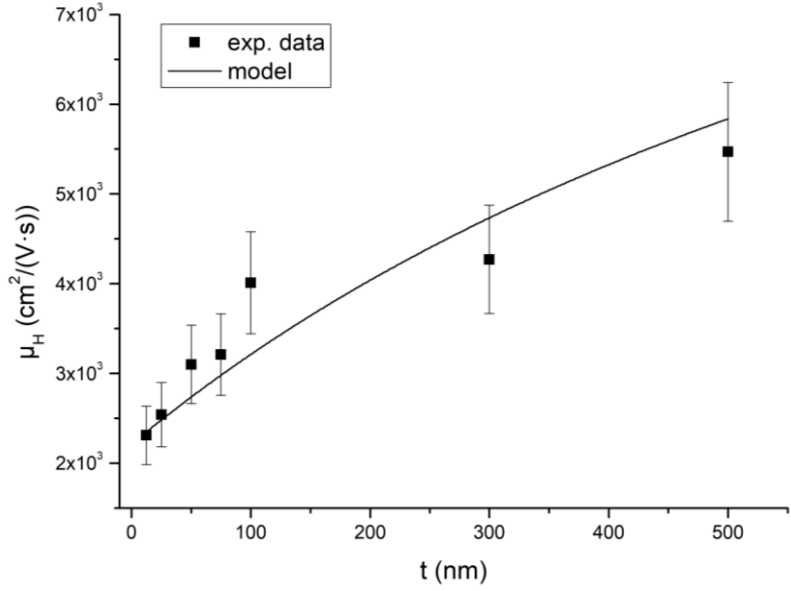

**Figure 8.** Hall mobility of samples A–G versus InAs thickness measured using the van der Pauw geometry at 300 K. The solid line was calculated from Equation (10).

Table 3. Summary of sheet resistance and Hall mobility of samples H-K measured at 300 K.

| Sample | t (nm) | $R_s$ (Ω/□) | $\mu_H$ (×$10^3$ cm$^2$/(V·s)) |
|---|---|---|---|
| H | 50 | 400 ± 40 | 4.1 ± 0.6 |
| I | 100 | 312 ± 31 | 4.8 ± 0.7 |
| J | 100 | 399 ± 40 | 2.3 ± 0.3 |
| K | 100 | 330 ± 33 | 2.1 ± 0.3 |

The behavior of the volume carrier concentration versus the InAs layer thickness for InAs layers grown on GaAs substrates was explained by the non-homogeneity of the carrier concentration along the growth direction [25]. Therefore, parallel conduction models were proposed to understand the transport properties of these structures. Since similar behavior was observed here for the InAs layers deposited on metamorphic buffers (Figure 6), we also interpreted our results with the inhomogeneity of carrier concentration along the growth direction. As reported in Ref. [26], three regions with different transport properties are expected for thick InAs layers grown on GaAs. We expect to find, for the InAs films on metamorphic buffers, two highly charged layers (layers 1 and 3) and one layer with a lower charge (layer 2), as shown in Figure 9.

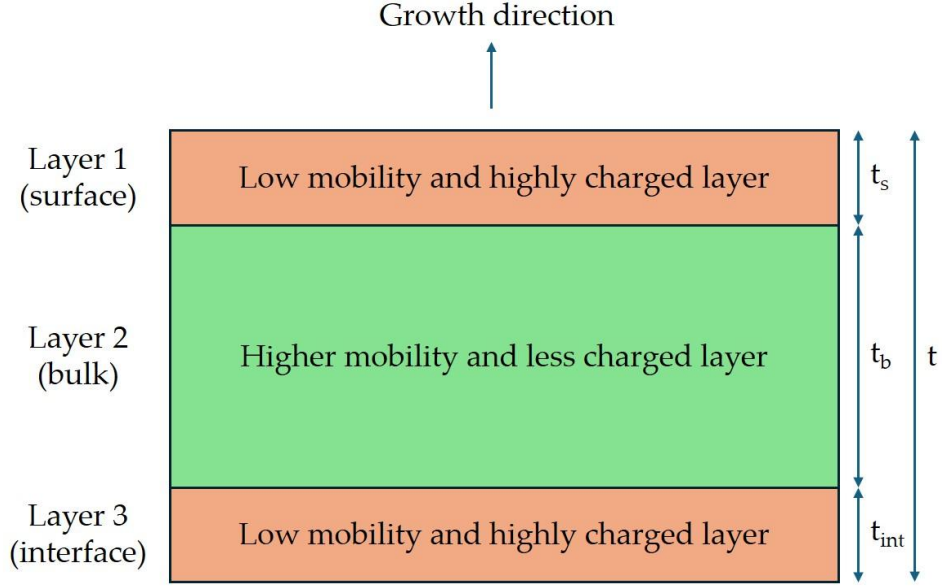

**Figure 9.** Schematic view of the three different conduction channels in the InAs layer.

- Layer 1 (surface layer): since the InAs surface Fermi level is pinned in the conduction band, which is bent at the InAs–vacuum interface, surface charge accumulation takes place, and the presence of a low-mobility surface layer is expected. A reasonable estimation for the thickness of this layer is the Debye length $L_D$. In particular, since the carrier concentration $n_b$ of a bulk-like near-intrinsic InAs crystal is of the order of $1 \times 10^{15}$ cm$^{-3}$ [27], we expect

$$L_D = \sqrt{\frac{\varepsilon_0 \kappa k_B T}{e^2 n_b}} \cong 130 \text{ nm} \qquad (5)$$

for T = 300 K and the relative permittivity of InAs $\kappa$ = 12.5.
- Layer 2 (bulk layer): the InAs layer corresponding to an InAs bulk crystal.
- Layer 3 (interface layer): a low-mobility and charged interface layer is expected to be present because of the nucleation of misfit dislocations in InAs layers that are thicker than the critical thickness of InAs on In$_{0.84}$Al$_{0.16}$As. Wolkenberg et al. [27] estimated the thickness of this highly defected layer for InAs on GaAs to be 1.8 nm. This value corresponds roughly to the critical thickness of InAs on GaAs. Here, we estimate this highly defected layer to be around 10 nm thick, which is the estimated critical thickness of InAs on this metamorphic buffer.

Let us call $t_s$, $t_b$, and $t_{int}$ the thickness of layer 1, layer 2, and layer 3, respectively, with $t = t_s + t_b + t_{int}$. Let $n_s$, $n_b$, and $n_{int}$ be the volume carrier concentration of layers 1, 2, and 3. Note that in this model, $t_s$ and $t_{int}$ are constant, while the thickness of layer 2 increases with increasing total film thickness. This model works well for relatively thick InAs layers grown on GaAs for which the Debye length $L_D$ (over which surface charge is screened) is smaller than the total layer thickness t. In the case of thin InAs layers on the metamorphic buffer, this can no longer be accurate. For t ≲ 100 nm, the value of $L_D$ is not negligible; thus, a separate bulk and surface charge region is not a good hypothesis. Therefore, we introduced a simplified two-parallel-channel conduction model.

This two-parallel-channel conduction model is made considering the following:

- One highly defected interface layer with thickness $t_{int}$ and carrier concentration $n_{int}$, which is layer 3 of the three-channel model.
- One layer that stands for the 'bulk–surface' transport properties with thickness ($t - t_{int}$) and carrier concentration $n_{bs}$.

Since $t_{int}$ is assumed to be ~10 nm, for each sample considered here, the film thickness is larger than $t_{int}$, and therefore we can reasonably assume $t_{int}$ to be constant for all films discussed here. The 'bulk–surface' layer is the layer with thickness (t − $t_{int}$) and carrier concentration $n_{bs}$. In this approach, $n_{bs}$ will be a weighted average of the surface $n_s$ and bulk $n_b$ carrier concentrations. For the thin InAs films, particularly for t ≲ $L_D$, we expect $n_{bs}$ to be close to the surface carrier concentration $n_s$ since the surface charge is not screened. For thicker 'bulk–surface' layers, the influence of $n_s$ will be smaller, and $n_{bs}$ is expected to approach the bulk InAs carrier concentration $n_b$. This means that we do not expect $n_{bs}$ to be constant as a function of t.

An expression for the Hall coefficient of a sample composed of two layers has been given by Petritz [28]:

$$R_H = \frac{t(R_{bs}\sigma_{bs}^2 t_{bs} + R_{int}\sigma_{int}^2 t_{int})}{(\sigma_{bs}t_{bs} + \sigma_{int}t_{int})^2}, \quad (6)$$

with $R_H = 1/ne$, $R_{bs} = 1/n_{bs}e$, $R_{int} = 1/n_{int}e$, and $\sigma_{bs}$ and $\sigma_{int}$ being the corresponding conductivities. We note that the same formula can be obtained from more general models for carrier concentration [29,30], neglecting higher-order terms in the magnetic field of the order of ($B^2$).

In going from the Hall coefficient to carrier concentration, we can rewrite Equation (6) as

$$\frac{1}{n} = \frac{t}{n_{bs}t_{bs} + n_{int}t_{int}} \times f(\sigma_{bs}, \sigma_{int}, n_{bs}, n_{int}, t_{bs}, t_{int}) \quad (7)$$

where $f$ is a factor that depends on thickness, carrier concentration, and conductivities of the two layers. Assuming reasonable parameters from [27,30], we calculate the value of $f$ to be of the order of one. Therefore, we set $f$ to 1 and obtain

$$n = n_{bs} + (n_{int} - n_{bs})\frac{t_{int}}{t} \quad (8)$$

The functional dependence reported in Equation (8) is the same as reported by Wolkenberg [19] for two parallel conduction channels. Equation (8) has been used to fit the thickness dependence of volume carrier concentration with $n_{bs}$ and ($n_{int} - n_{bs}$)·$t_{int}$ as fitting parameters. However, since the fit assumes a constant $n_{bs}$, the obtained value of $n_{bs}$ will be an average value of $n_{bs}$ through the dataset. The fitting procedure is performed considering the uncertainty of the single data points. The best fit to the experimental data is reported in Figure 6. From the fitting parameters, we obtain $n_{bs}$ = (5.1 ± 1.4) × $10^{16}$ cm$^{-3}$ and ($n_{int}-n_{bs}$)·$t_{int}$ = (4.9 ± 0.3) × $10^{12}$ cm$^{-2}$. As we can see in Figure 6, the model is in good agreement with the experimental data. Moreover, since $t_{int}$ ∼ 10 nm, from the value of ($n_{int} - n_{bs}$)·$t_{int}$ we can estimate $n_{int} \simeq 4.8 \times 10^{18}$ cm$^{-3}$. This value is ≃10 times smaller than the value reported in Ref. [27], ∼2 × $10^{19}$ cm$^{-3}$. This difference can easily be explained in terms of the higher dislocation density expected for InAs on GaAs with respect to InAs on a metamorphic buffer.

Samples H and I are measured to obtain an estimate of the role of the surface layer. Since these two samples are capped, we expect that surface charge accumulation is strongly reduced. We notice that the reduction in the sheet carrier concentration of samples H and I compared to the not-capped samples are (2.8 ± 0.8) × $10^{12}$ cm$^{-2}$ and (1.6 ± 0.7) × $10^{12}$ cm$^{-2}$, respectively. This means that the reduction in the sheet carrier concentration of samples H and I is comparable within their error, consistent with the assumption of a constant reduction in surface charge density. We expect the reduction in carrier concentration in the capped samples with respect to the not-capped samples to be of the order of the surface layer sheet carrier concentration. Therefore, we estimate the surface layer sheet carrier concentration ($n_s t_s$) as the average reduction in carrier concentration measured for samples H and I, resulting in $n_s t_s$ = (2.2 ± 0.5) × $10^{12}$ cm$^{-2}$. In addition, from the fit of Figure 6 with Equation (8), we obtained an interface layer sheet carrier concentration ($n_{int}t_{int}$) of (4.8 ± 0.3) × $10^{12}$ cm$^{-2}$. Therefore, the sum of the interface and surface sheet carrier concentrations is ($n_s t_s$ + $n_{int}t_{int}$) = (7.0 ± 0.6) × $10^{12}$ cm$^{-2}$. The measured value of the sheet carrier concentration of the 50 nm and 100 nm thick InAs samples grown on metamorphic buffers are (6.6 ± 0.7) × $10^{12}$ cm$^{-2}$ and (5.8 ± 0.6) × $10^{12}$ cm$^{-2}$, respectively. These two values are comparable to ($n_s t_s$ + $n_{int}t_{int}$) within the experimental errors. Therefore, we conclude that for thin samples, most of the charge is due to the surface and interface layers. Furthermore, we note that for the 12.5 nm and 25 nm thick InAs films (samples A and B), the value of $N_s$ is $(3.4 \pm 0.4) \times 10^{12}$ cm$^{-2}$ and $(5.4 \pm 0.5) \times 10^{12}$ cm$^{-2}$. These two values are not comparable with the estimated sum of the interface and surface sheet carrier concentrations. This result highlights the limitation of our model for InAs thickness of the order of $t_{int}$. Indeed, for InAs thickness of the order of the critical thickness, a lower number of dislocations is expected, resulting in a lower interface sheet carrier concentration ($n_{int}t_{int}$) and thus a less-charged InAs layer.

The model of Equation (8) can also be developed to fit the layer conductivity. Calling σ the conductivity, we can write [27–30]

$$\sigma = \sigma_{bs} + (\sigma_{int} - \sigma_{bs})\frac{t_{int}}{t} \qquad (9)$$

From the fit of the conductivity, ($\sigma = 1/(R_s \cdot t)$), $\sigma_{bs} = (1.1 \pm 0.3) \times 10^2$ Ω$^{-1}$·cm$^{-1}$ and $(\sigma_{int} - \sigma_{bs}) \cdot t_{int} = (1.7 \pm 0.3) \times 10^4$ Ω$^{-1}$·cm$^{-1}$ nm are derived. As a consistency test, we calculated the factor $f$ of Equation (7) for the obtained conductivities and carrier concentrations, resulting in $1 < f < 2$ for every InAs layer thickness.

Since $\mu_H = \sigma/(ne)$, we finally obtain

$$\mu_H = \frac{(\sigma_{int} - \sigma_{bs})t_{int} + \sigma_{bs}t}{[(n_{int} - n_{bs})t_{int} + n_{bs}t]e} \qquad (10)$$

Figure 8 plots the Hall mobility using Equation (10) using the best-fit parameters obtained from Equations (8) and (9).

## 4. Conclusions

In this work, InAs layers with thickness t ranging from 12.5 nm to 500 nm were grown by solid-source MBE on an In$_x$Al$_{1-x}$As double slope step-graded buffer. HR-XRD was employed to analyze the samples. The strain relaxation of the InAs layers on the metamorphic buffers agrees with the model by Dunstan [22] predicting a 1/t relaxation. The FWHM was studied as a function of the InAs thickness. The measured FWHM agrees with a 1/t dependence, consistent with a thickness broadening of the InAs peak. We performed Hall effect measurements on these samples at 300 K in the van der Pauw configuration. Since no accurate theory of carrier concentration and electron mobility (like a Schrödinger–Poisson solver) is available at present, we developed a simple two-parallel conduction model composed of a highly defected interfacial layer and a 'bulk–surface' InAs layer. The model is in good agreement with the experimental data. To quantify the surface charge contribution to the transport data, two samples consisting of a 50 and 100 nm thick InAs layer on a metamorphic buffer were grown with an In$_{0.84}$Al$_{0.16}$As 5 nm thick cap layer on top of it, which suppresses the surface charge accumulation. The transport measurements of these two samples show that for the thin layers (t ≲ 100 nm), most of the charge in the InAs film is due to the surface and interface charge regions. However, for the 12.5 nm and 25 nm thick InAs epilayers, the values of $N_s$ are lower than the estimated sum of the interface and surface sheet carrier concentration. This result highlights the limitation of our model for InAs thickness of the order of $t_{int}$. Indeed, for InAs thickness of the order of the critical thickness, a lower number of dislocations is expected, resulting in a lower interface sheet carrier concentration and, thus, a less-charged InAs layer.

In addition, InAs films were grown directly on GaAs at two different growth temperatures. We found that both samples have a higher carrier concentration and lower mobility with respect to the InAs on the metamorphic buffers. Moreover, we conclude that a higher crystal quality is obtained for the InAs film grown on the metamorphic buffer compared to that directly grown on the GaAs substrate.

Understanding the carrier location in the InAs epilayers grown on metamorphic buffers will allow the design and fabrication of surface-exposed classic and superconducting field-effect transistors with superior gate-tunable electrical properties.

**Acknowledgments:** We thank Robert Andrei Sorodoc for measuring the optical microscopy images. This research was funded by EU's Horizon 2020 Research and Innovation Framework Program under Grant No. 964398 (SUPERGATE), No. 101057977 (SPECTRUM), and the PNRR MUR project PE0000023-NQSTI.